# Epitaxy and Magneto-transport properties of the diluted magnetic semiconductor p- Be$_{(1-x)}$Mn$_x$Te

L. Hansen[1], D. Ferrand[2], G. Richter, M. Thierley, V. Hock, N. Schwarz, G. Reuscher, G. Schmidt, A. Waag[*], L.W. Molenkamp

Physikalisches Institut der Universität Würzburg, EP III, 97074 Würzburg, Germany

[*]Abteilung Halbleiterphysik, Universität Ulm, 89081 Ulm, Germany

Abstract

We report on the MBE-growth and magnetotransport properties of p-type BeMnTe, a new ferromagnetic diluted magnetic semiconductor. BeMnTe thin film structures can be grown almost lattice matched to GaAs for Mn concentrations up to 10% using solid source MBE. A high p-type doping with nitrogen can be achieved by using an RF-plasma source. BeMnTe and BeTe layers have been characterized by magneto-transport measurements. At low temperatures, the BeMnTe samples exhibit a large anomalous Hall effect. A hysteresis in the anomalous Hall effect appears below 2.5K in the most heavily doped sample, which indicates the occurrence of a ferromagnetic phase.

---

[1] Electronic mail: hansen@physik.uni-wuerzburg.de
[2] Present address: Laboratoire de Spectometrie Physique, 38402 Grenoble cedex, France

Diluted magnetic semiconductors (DMS) have attracted more and more attention during the last years, because such magnetic epilayers can be conveniently integrated into semiconductor heterostructures to combine the properties of both, magnetic materials and semiconductors. First experiments have shown the possibility to inject spin-polarized electrons aligned by a paramagnetic II–VI semiconductor layer in an external magnetic field into an (Al)GaAs-LED [1]. A next step could be a combination of a ferromagnetic material with a semiconductor to avoid the necessity of applying an external magnetic field for the orientation of electron or hole spins. Since it seems to be very difficult to inject spin-aligned carriers from a ferromagnetic metal into a semiconductor [2], it is essential to look for new ferromagnetic semiconductors with a smaller conductivity than metals, which show a complete spin polarization and which can be grown onto the common non-magnetic semiconducting materials as e.g. GaAs.

Hole-induced ferromagnetism has been observed already in III-V DMS layers of GaMnAs [3] and InMnAs [4,5] as well as in II-VI heterostructures (CdMnTe QW [6], ZnMnTe layers [7]). The advantage of II-VI compounds is to control the localized spins and the hole concentration independently, which makes them particularly attractive for fundamental studies. In this paper we report on the observation of ferromagnetism of highly p-type doped BeMnTe, which can be grown almost lattice matched onto GaAs.

Growth was carried out in a multi chamber Riber 2300 MBE system by solid source MBE on semi-insulating (001) GaAs substrates using elemental Be, Mn and Te effusion cells. To improve the quality of the substrate surfac a 500 nm thick GaAs buffer layer was grown in a separate III/V-MBE-chamber. The sample was then transferred under UHV conditions to the II/VI-chamber.

Residual Selenium also present in the II/VI MBE-chamber is known to react with GaAs resulting in an amorphous surface layer [8]. To prevent this surface degradation, immediately after inserting the wafer into the II/VI MBE-chamber a BeTe buffer of 5 ML thickness was grown at a substrate temperature of 350°C which has been suggested by Fischer et al. for the growth of ZnSe on GaAs [9]. Before incorporating Mn the substrate temperature was lowered to 300°C. Growth was monitored in situ by RHEED. P-type doping with nitrogen was accomplished by a RF-Plasma source with an input rf power of 450 W. The total optical emission intensity from the rf plasma could be monitored using a photo diode. The residual nitrogen pressure remained in the $10^{-6}$ Torr range during growth. To protect the Be(Mn)Te layers from oxidation during post growth processing, a 5nm thin p-type doped ZnSe cap layer was grown on top of each sample before an in-situ Au contact was evaporated. 5 nm of p-ZnSe are still thin enough to provide an ohmic contact to the p-BeTe [10].

To determine the exact Mn-content and to evaluate the crystal quality X-ray-diffraction measurements have been done using a five-crystal high-resolution X-ray diffractometer (HRXRD). Net acceptor concentration ($n_a - n_d$) and (anomalous) Hall effect were measured by etching a Hall bar in a six point geometry. Transport experiments were done at magnetic fields from 0 to 7 Tesla from room-temperature down to 1.6 K.

All growth was carried out under Te-rich conditions which is indicated by a (2x1) reconstruction visible by RHEED. Be-rich growth leads to a change from a (2x1) RHEED pattern towards (4x1) [11]. For BeTe, a Be-rich growth results in a poor crystalline quality due to the formation of Be-clusters. On the other hand, a too high Te-flux hinders the incorporation of nitrogen acceptors [11]. As an optimum we chose a beam equivalent pressure (BEP) (Te/Be) = 10.

A first estimate of the Mn-content in the epilayer can be extracted from the growth rates determined from RHEED oscillations. Fig. 1 shows pronounced RHEED oscillations. The growth-rate of the BeTe-buffer is determined as 0.24 ML/s whereas additional Mn leads to a rate of 0.28 ML/s. Exact values can be obtained by HRXRD applying Vegard´s law. The sample shown in Fig. 1 contains according to HRXRD 11.9% Mn, less than one would have estimated on the basis of RHEED oscillations. We have not investigated yet, whether the sticking coefficient of Mn on a BeTe surface is already diminished from 1 in the optimum temperature range of 300 – 350°C for BeTe-growth. In any case, it is apparently not necessary for substitutional Mn incorporation to lower the substrate temperature to a regime which would degrade the quality of the epi-layer (below 300°C). This is in contrast to GaMnAs where one of the main problems is the low growth temperature which leads to a reduced crystalline quality of the GaAs-matrix [12].

By measuring the (004), (115) and (-1-15) X-ray reflection one can obtain the strain state of the epilayers. Fig. 2 shows a HRXRD $\omega$-$2\theta$ scan of a 130 nm thin almost fully strained (93 %) layer containing 7.25 % of Mn. Pronounced thickness fringes indicate the good quality of the sample, the BeMnTe peak itself is broadened by approximately 110 arc sec, which is due to the intrisnic broadening of the reflection because of the small thickness of the layer. Whereas BeTe has a slightly smaller lattice constant (5.627 Å [13]) than GaAs (5.653 Å), MnTe in its zincblende phase has a much higher (6.337 Å) [14]. Therefore, the incorporation of small amounts of Mn rather improves than diminishes the quality of Be(Mn)Te epilayers. Several samples were grown containing 0 to 18 % of Mn. For a thickness of less than 100 nm even for 12% Mn good crystalline quality is achieved (the samples were still fully strained), whereas for 17.7 % Mn content growth turns immediately towards 3 D due to the large mismatch to the GaAs substrate. Most investigated samples contained up to 10 % Mn which led to good ferromagnetic properties (see below).

The maximum p-type dopability of BeTe has been reported to be up to $10^{20}$ cm$^{-3}$ [15], a value which is comparable with the doping levels obtained in Zn(Mn)Te [7].

Particularly this high p-type dopability focussed our interest on BeMnTe. Using a RKKY model, the ferromagnetic coupling in these tellurium compounds is mediated by the holes and overcompensates the superexchange anti-ferromagnetic interaction of the $Mn^{2+}$ spins [7]. Crucial for a high p-doping is the VI / II flux ratio which should be approximately 10. Too high Te-flux leads to a decrease in built in of nitrogen dopant [11]. A second important factor is the efficiency of the utilized plasma source. For the samples described in this paper we used a beam-plate with only a single aperture of approximately 0.3 mm in diameter, leading to an acceptor concentration of $10^{19}$ $cm^{-3}$ (sample 1476) in BeTe at a typical growth rate of 0.1 ML/s. Table 1 shows an overview of a set of 3 Be(Mn)Te-films grown in a row with the same VI / II ratio, growth rate and plasma cell intensity (1476-78). Incorporation of 4.2 and 6.9 % Mn, respectively, did not effect the electronic properties (samples 1477 and 1478). All 3 samples have the same doping level of $10^{19}$ $cm^{-3}$ and typical hole mobilities of around 30 $cm^2$/Vs. Also crystalline qualities were good resulting e.g. in pronounced thickness fringes in HRXRD. A second set (1521 and 1526) showed similar properties. To increase the acceptor concentration, we decreased the growth rate by almost a factor of 10. This results in very long growth times of several hours for only 100 nm thin samples and rather poor crystalline quality, also visible in a decline in hole mobility by a factor of 3 (sample 1567). No RHEED-oscillations could be observed at this low growth rate. Nevertheless, the measurable doping level could be increased by a factor of more than 3.

For an investigation of the transport properties of p-BeMnTe, we first characterized 2 samples, one containing 6.5 % of Mn (1521), the other one was non magnetic (1526). From room temperature Hall measurements for both samples a p-type doping of about 1 x $10^{19}$ $cm^{-3}$ could be deduced. The non magnetic p-BeTe reference sample exhibited the expected linear Hall voltage down to 1.7 K and stayed metallic in the whole temperature range. In contrast, the p-$Be_{0.935}Mn_{0.065}$Te sample exhibited a metal-insulator transition and a typical huge negative magnetoresistance in the longitudinal transport at temperatures below 10 K and magnetic fields below 3 Tesla. Measurements of the transverse resistance showed an anomalous Hall effect, though no hysteretic behavior on the magnetic field could be observed. The former behavior has been reported previously for p-ZnMnTe and been interpreted as carrier localization as bound magnetic polarons [7], a first hint on the interaction between the Mn-ions mediated by holes. As has been observed earlier in the case of ZnMnTe, a slight increase in doping can drastically change the situation. In fact, the higher doped sample 1567 ($p=3.4 \times 10^{19}$ cm$^{-1}$) stays at B=0 in the metallic phase down to 1.6 K. Fig. 3 shows a hysteresis in the anomalous Hall effect below 3 K, indicating a ferromagnetic transition at $T_c = 2.5$ K. Further analysis of the magnetic properties e.g. by SQUID measurements is under work.

In this paper we discussed the properties of the new DMS BeMnTe which can be grown by MBE on GaAs. Details on growth and p-type doping with a nitrogen plasma source are given. For a highly p-type doped sample ($3.4 \times 10^{19}$ cm$^{-3}$) containing 10% of Mn we have shown a hysteretic behaviour of the anamolous Hall effect, indicating a ferromagnetic state below 3 K. We assume that a further increase in doping with an optimized plasma source should lead to an increase in $T_c$. With the possiblity of incorporating more than 10% of Mn ions into BeTe this new material offers a huge field for investigations of the involved magnetic mechanisms in DMS, such as antiferromagnetic coupling of the Mn spins and the ferromagnetic interaction mediated by the holes.

This work has been supported by the Deutsche Forschungsgemeinschaft through SFB 410.

Table 1: Investigated samples, electric properties deduced from transport measurements at RT.

| sample No. | thickness [nm] | x(Mn) [%] | hole density [cm$^{-3}$] | conductivity [S cm$^{-1}$] | mobility [cm$^2$/Vs] |
|---|---|---|---|---|---|
| 1476 | 75 | 0 | 10$^{19}$ | 54 | 34 |
| 1477 | 60 | 4.2 | 10$^{19}$ | 46 | 29 |
| 1478 | 75 | 6.9 | 10$^{19}$ | 53 | 33 |
| 1521 | 118 | 6.5 | 10$^{19}$ | 46 | 34 |
| 1526 | 143 | 0 | 1.4 10$^{19}$ | 45 | 45 |
| 1567 | 60 | 10 | 3.4 10$^{19}$ | 57 | 10.5 |

Figure captions

Fig. 1: RHEED oscillations obtained during growth of a 5 ML BeTe buffer layer on GaAs, followed by the growth of BeMnTe.

Fig. 2: HRXRD rocking curves of (004), (115 and (-1-15)-reflexes of 130nm Be$_{0.93}$Mn$_{0.07}$Te.

Fig. 3: Hall resistance vs. magnetic field for p-BeMnTe (sample 1567) containing 10 % Mn, grown at 1/10 of the usual growth rate, T= 1.6 –4.2 K.

Hansen et al., Fig.1:

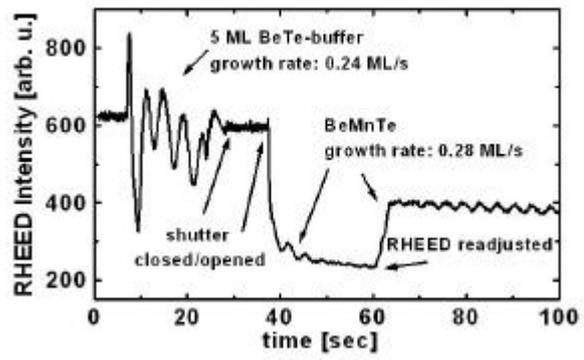

Hansen et al., Fig.2:

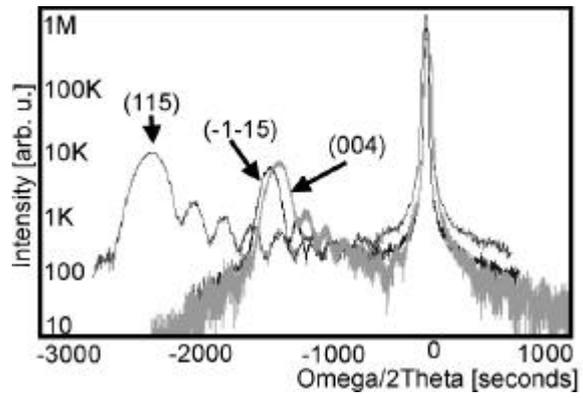

Hansen et al., Fig.3:

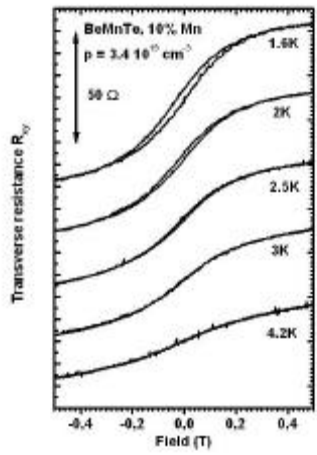

[15] H.-J. Lugauer, Th. Litz, F. Fischer, A. Waag, T. Gerhard, U. Zehnder, W. Ossau, G. Landwehr, J. Crystal Growth **175/176**, 619 (1997).